\documentclass[useAMS,usenatbib]{aa}
\usepackage{txfonts}
\usepackage{natbib}
\usepackage{longtable}
\usepackage[dvips]{graphicx}
\author{R. Voss\inst{1}, P. Martin\inst{2}, R. Diehl\inst{3}, J.S. Vink\inst{4}, D.H. Hartmann\inst{5}, T. Preibisch\inst{6}}
\institute{Department of Astrophysics/IMAPP, Radboud University Nijmegen, PO Box 9010, NL-6500 GL Nijmegen, the Netherlands \and Institut de Plan\'etologie 
et d'Astrophysique de Grenoble, BP 53, 38041, Grenoble cedex 9, France \and
Max-Planck-Institut f\"ur extraterrestrische Physik, Giessenbachstrasse, 
D-85748, Garching, Germany \and Armagh Observatory, College Hill, Armagh, BT61 9DG, Northern Ireland, UK \and Department of Physics and Astronomy, Clemson University, Kinard Lab of Physics, Clemson, SC 29634-0978 \and Universit\"ats-
Sternwarte M\"unchen, Ludwig-Maximilians-Universit\"at, Scheinerstr. 1, 
81679, M\"unchen, Germany}

\authorrunning{Voss et al.}
\titlerunning{Massive stars and supernovae in Carina}
\title{Energetic feedback and $^{26}$Al from massive stars and 
their supernovae in the Carina region}
\abstract
{}
{We study the populations of massive stars in the Carina region and
their energetic feedback and ejection of $^{26}$Al.
}
{
We present a census of the stellar populations in young stellar clusters 
within a few degrees of the Carina Nebula. For each star we estimate 
the mass, based on the spectral type and the host cluster age. We use
population synthesis to calculate the energetic feedback and ejection
of $^{26}$Al from the winds of the massive stars and their supernova 
explosions. We use 7 years of INTEGRAL observations to measure the 
$^{26}$Al signal from the region.
}
{
The INTEGRAL $^{26}$Al signal is not significant with a best-fit
value of $\sim1.5\pm1.0\times10^{-5}$ ph cm$^{-2}$ s$^{-1}$,
approximately half of the published Compton Gamma Ray Observatory (CGRO) 
result, but in
agreement with the latest CGRO estimates.
Our analysis of the stellar populations in the young clusters
leads to an expected signal of $\sim$half the observed value,
but the results are consistent within 2$\sigma$. 
We find that the fraction of $^{26}$Al
ejected in Wolf-Rayet winds is high,
and the observed signal is unlikely to be caused by $^{26}$Al ejected in
supernovae alone, indicating a strong wind ejection of $^{26}$Al.
Due to the lack of prominent O stars, regions with ages $\gtrsim$10 Myr
are often neglected in studies of OB associations. We find that in the
Carina region such clusters contribute significantly
to the stellar mass and the energetics of the region. 
}
{}

\begin{document}
\maketitle
\section{introduction}
Feedback from massive stars plays a crucial role in the formation
of stars and in shaping the surrounding inter-stellar medium (ISM).
We developed a new population synthesis tool to study the feedback
from populations of massive stars in OB associations \citep{Voss-popsyn}.
In \citet{Voss-orion} we applied the population synthesis to the nearby 
Orion region and found good agreement with observations of the region. 
However, the population of massive stars in Orion is not large enough
to provide strong constraints on the feedback models. For this it is
necessary to study a larger population including very high mass
stars ($\sim$100$M_{\odot}$).

The Carina region hosts a large population of very young massive
stars at a distance of 2.3$\pm0.1$ kpc \citep{Allen1993,Walborn1995,Smith2002}
including 72 of spectral type O \citep{Smith2006,Cappa2008},
6 Wolf-Rayet (WR) stars \citep{vanderHucht2001}, one luminous blue
variable (LBV;$\eta$ Carinae) and three evolved red supergiants
\citep{Feinstein1980,Feinstein1981}. The total stellar mass
is estimated to be $\sim3.7\times10^4 M_{\odot}$ \citep{Preibisch2011b},
and the total mass of the surrounding gas and dust is
$\sim2.8\times10^{5} M_{\odot}$ \citep{Preibisch2011a}.
The majority of the young
stars reside in the Carina nebula \citep[summarized in][]{Smith2006},
but the surrounding region hosts a large number of smaller open
clusters with a wider range of stellar ages. 
While the stellar populations have been studied in the
most prominent of these \citep[NGC3293;NGC 3324][]{Evans2005,Cappa2008},
many of the smaller clusters were never investigated in detail.

The content of massive stars in the Carina region is intermediate
between the population of relatively well-studied small star-forming 
regions, such as Orion and Sco-Cen, and more distant superclusters,
where single regions such as 30 Doradus, hosting $\sim$1000 O-stars,
can affect the energetics and chemistry of their host galaxies. 
Therefore the Carina region is useful for the study of massive stars,
feedback from massive stars, and important for the understanding of how
the feedback mechanisms scale with the size/mass of the region.
The population is similar to the Cygnus OB2 association hosting
80 \citep{Hanson2003} stars of type O, which was the goal of a
recent study similar to ours \citep{Martin2009,Martin2010}. 

The radioactive isotope $^{26}$Al is ejected from massive stars
through their winds and supernova explosions \citep{Prantzos1996}. 
It is therefore
intimately related to the energy feedback from massive stars.
It has a mean lifetime of $\sim1$ Myr and is traced by the
$\gamma$-ray decay line at 1808.63 keV, observable by $\gamma$-ray
observatories such as the COMPTEL instrument aboard the CGRO
and the SPI instrument aboard INTEGRAL. 

In the light of the uncertainties in the modelling of massive
stars, it is necessary to perform multi-wavelength consistency checks, 
encompassing the different aspects of feedback provided by massive star 
clusters. In this paper we discuss the ejection of $^{26}$Al and the 
injection of energy into the ISM from the massive star population in
the Carina region. This is complementary to the modelling and observations
of the energy and UV balance studied by \citet{Smith2006,Smith2007}.
The comparison between the theoretical models
and the observables are important, both for understanding crucial
parts of stellar evolution, in particular mass loss rates,
nucleosynthesis and supernova explosions, and for understanding
issues related to the ISM, such as star-formation and feedback
mechanisms.
A new generation of stellar models \citep{Meynet2005,Palacios2005,Limongi2006}, 
improved observations of stellar populations in star-forming regions 
and the advent of the INTEGRAL observatory providing new $^{26}$Al 
observations, has allowed progress on the subject:
\citet{Voss-orion} studied the variations
between different models of massive stars, in particular the
effects of rotation and the strength of wind mass-loss 
on the the radio-active tracers and the energetics of star-forming
regions. The individual nearby star-forming regions
Sco-Cen \citep{Diehl2010}, Orion \citep{Voss-orion} and Cygnus
\citep{Martin2009,Martin2010} have been studied in detail and good
agreement has been found between theory and the observations.

\section{Observations of $^{26}$Al from the Carina region}
\label{sect:observations}
\subsection{CGRO results}
The $^{26}$Al signal from the Carina region was discussed by
\citet{Knoedlseder1996}, based on a measured flux of
3.2$\times10^{-5}$ ph s$^{-1}$ cm$^{-2}$ (with statistical and
systematic uncertainties of 0.8$\times10^{-5}$ and 0.1$\times10^{-5}$,
respectively). Their analysis showed that the signal is seen from
within 2 degrees of the Carina Nebula direction only, although
an origin from a larger region extending up to 6 degrees would
also be consistent with the measurements. A subsequent analysis
of the full CGRO database revised flux values down to
1.1--2.2$\times10^{-5}$ ph s$^{-1}$ cm$^{-2}$ \citep[with $\sim20$\%
statistical error][]{Pluschke2001}. This signal
corresponds to 0.005--0.010 $M_{\odot}$ of $^{26}$Al at the distance
of the Carina Nebula. An origin from foreground stellar groups
and from AGB stars is help implausible, and background groups
would have to be exceptionally active to explain the flux
\citep[see][for a discussion]{Knoedlseder1996}. Here we use
INTEGRAL observations for a new measurement of $^{26}$Al
from the Carina region, and use a population synthesis approach
\citep{Voss-popsyn} to compare to expectations.

\subsection{INTEGRAL data analysis}
We used INTEGRAL data taken between revolution 19 and 855 of 
the satellite. This corresponds to a total effective exposure of 
17.6Ms on Carina, defined as a circular region with radius of 
40$^\circ$ centred on $(l,b)=(285.0^\circ, 0.0^\circ)$. The methodology 
developed in \citet{Martin2009} to extract the 1809keV signal from 
Cygnus was again used for the present study. The reader is referred 
to the latter work for more details about data preparation, instrumental 
background modelling, and other technical aspects. We searched for 
emission in the 1806--1812keV band over the 245$^\circ$--325$^\circ$ longitude 
range, using two different methods. 

\begin{figure}
\includegraphics[width=0.9\columnwidth,angle=0]{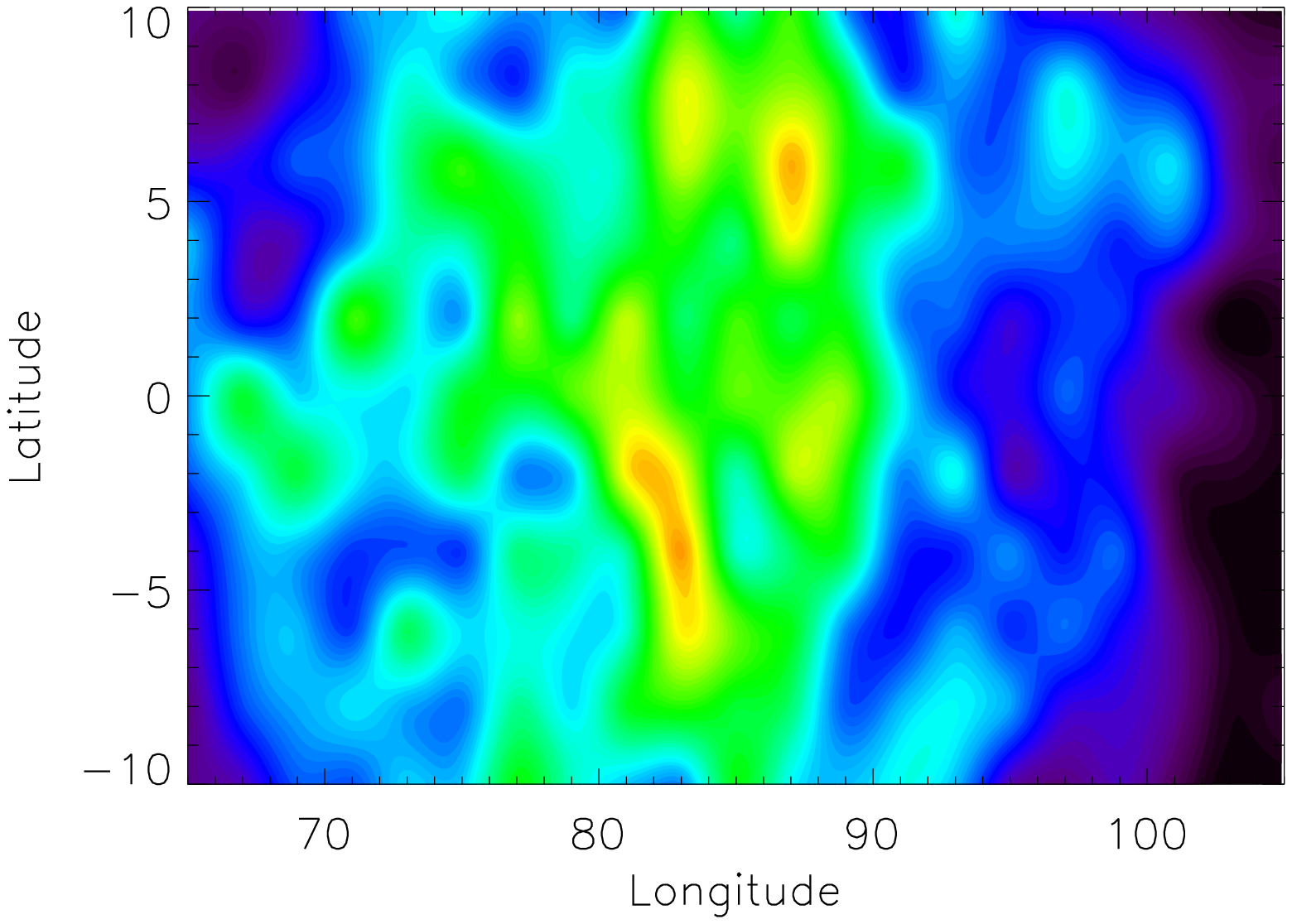}
\includegraphics[width=0.9\columnwidth,angle=0]{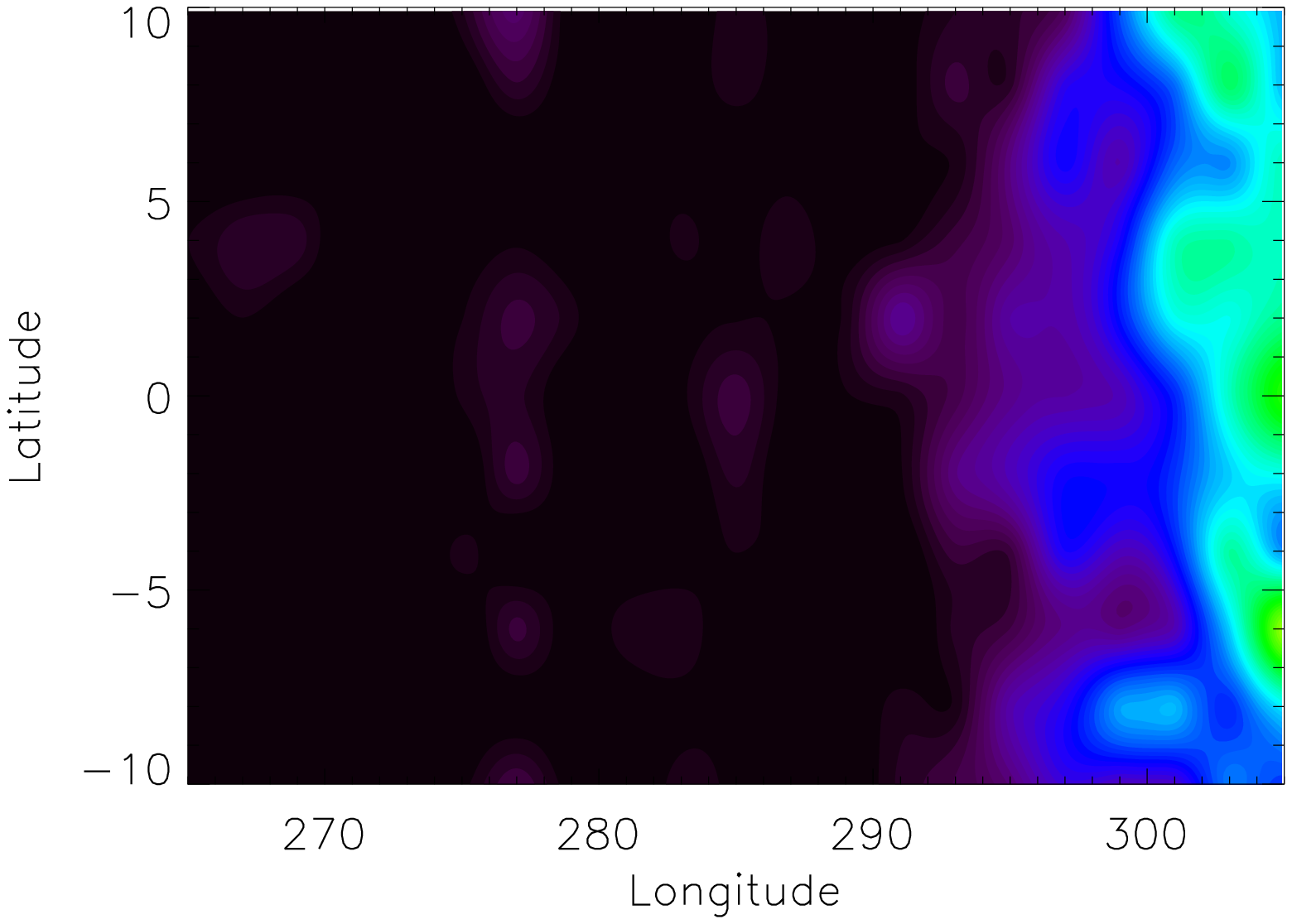}
\caption{Maximum Likelihood Ratio (MLR) maps built from a point-source 
scan of the regions of Cygnus (top panel) and Carina (bottom panel).
 The color coding ranges from MLR$\sim$0 (black) to MLR$\sim$40 (red).}
\label{fig:mlrmap}
\end{figure}

The first was a point-source scan, where we try to account for the 
data by the fitting of an instrumental background model and a sky 
model consisting of a single point-source. The operation is repeated 
for a grid of positions covering the whole Carina region.  
From this analysis, however, no significant excess emission 
was detected in the Carina region around $(l,b)=(287.0^\circ, 0.0^\circ)$ 
where the main massive star clusters are located. Significant emission 
appears beyond $l=305^\circ$, as we move towards the Galactic ridge, 
which dominates the allsky 1809keV emission. As a comparison, we did 
the same for the Cygnus region and found strong emission around 
$(l,b)=(80.0^\circ, 0.0^\circ)$, consistent with the dedicated analysis 
exposed in \citet{Martin2009} (while the exposure is similar for both 
regions, Cygnus hosts almost twice the number of O-stars, and is nearer). 
Fig. \ref{fig:mlrmap} shows the maximum likelihood ratio maps obtained 
for the Carina and Cygnus regions.

We then determined more quantitatively the 1809keV flux from the Carina 
region by fitting to the data extended sky models that describe 
better the expected $^{26}$Al decay emission from a conglomerate of 
star clusters. We used 2D Gaussian intensity distribution of various sizes 
and tried two different positions for these: the first is 
$(l,b)=(287.6^\circ,- 0.6^\circ)$, which is the position of Trumpler 16, 
the richest cluster in terms of O stars, and the second is 
$(l,b)=(285.8^\circ, 0.1^\circ)$, which is the position of NGC3293, 
which is the cluster with the largest number of past supernovae 
(as extrapolated from its present-day IMF, see below). In that way, 
we tested two scenarios: one in which the $^{26}$Al content of the 
Carina region is assumed to be fed mostly by the WR-winds of 
present-day massive stars, and one in which it is assumed to result 
mostly from past supernovae. Fluxes in the range 1.0--1.5 
$\times 10^{-5}$ ph s$^{-1}$ cm$^{-2}$ are obtained, with typical 
statistical uncertainties of order 1.0 $\times 10^{-5}$ ph s$^{-1}$ cm$^{-2}$. 
Combining information from 
both CGRO and INTEGRAL apparently points to a 1809keV flux 
from the Carina region, in the range 1--2 $\times 10^{-5}$ ph s$^{-1}$ 
cm$^{-2}$, corresponding to a $^{26}$Al mass of $4-9\times10^{-3}M_{\odot}$. 

\section{Analysis of the stellar populations}
\label{sect:stars}
\begin{table}
\begin{center}
\caption{The open clusters in our study.}
\label{tab:regs}
\begin{tabular}{lccc}
\hline\hline
Cluster & Number of O stars   & Age &  Distance \\ 
&& Myr & kpc\\
\hline
Bochum 10 & 1 & 7 & 2.3\\
Bochum 11 & 5 & 2 & 2.3\\
Loden 153 & 1 & 5.5 & 2.7\\
NGC 3293 & 0 & 10 & 2.7\\
NGC 3324 & 3 & 2.5 & 3.0\\
Trumpler 14 & 9 & 1 & 2.3\\
Trumpler 15 & 6 & 6 & 2.3\\
Trumpler 16 & 43 & 2.5 & 2.3\\
\hline
\end{tabular}\\
\end{center}
\end{table}

To compute the amount of $^{26}$Al in the observed region it is
necessary to understand the stellar content of the region, and how
star formation developed in the last $\sim10$ Myr. In our analysis
we include the region considered by \citet{Knoedlseder1996} to be
the emitting region, that is a region approximately centered on the
Carina Nebula, with a radius of $\sim2.8^\circ$. This includes the
Carina Nebula itself, but also some clusters that are distant
enough from this to not be linked directly to this by dynamics.
In our main analysis we only consider clusters with a distance 
below $\sim3$ kpc, but a possible contribution from the background 
will be discussed in Sect. \ref{sect:26al}.

The population of massive stars in clusters belonging to the
Carina Nebula was presented in \citet{Smith2006}, and we adopt
their stellar classifications and cluster ages (noting that
ages of such young clusters are always very uncertain)
and assume that the study is complete for
the O stars. In addition we include the three supergiants
listed in \citet{Feinstein1980,Feinstein1981}.
For NGC 3324 and NGC 3293 we used \citet{Cappa2008}
and \citet{Evans2005}, respectively, with the addition of
a red supergiant from \citet{Carraro2001}.
Other clusters were found
to be too old or small to contribute significantly to the observed
$^{26}$Al signal e.g. Loden 165, NGC 3114, VdB Hagen 99, Bochum 9
\citep{Carraro2001,Patat2001}. Additional WR stars that are
found outside the main clusters were taken from \citet{vanderHucht2001}.
Table \ref{tab:regs} summarizes basic information about the clusters
considered. Recent studies have shown that there is a significant
population of OB stars obscured by dust from the Carina region itself
\citep[e.g.][]{Povich2011}.
We will discuss this population in Sect. \ref{sect:tvo}.

\begin{figure*}
\resizebox{\hsize}{!}{\includegraphics[angle=0]{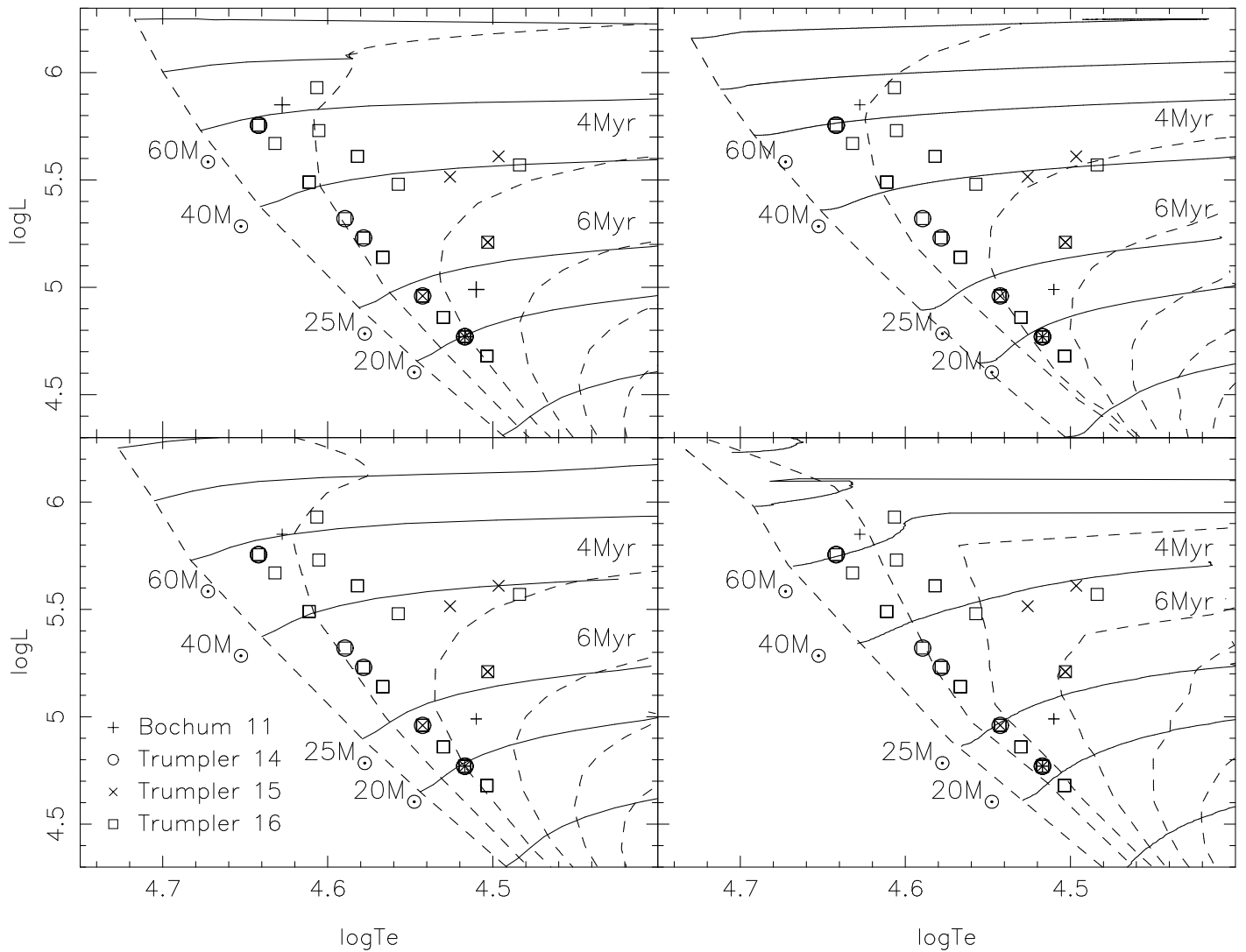}}
\caption{The O stars in the 4 groups richest in O stars.
The solid lines corresponds to stellar tracks, and the dashed lines
are isochrones with a separation of 2 Myr.
The stellar models are from \citet{Meynet1997} (upper left),
\citet{Limongi2006} (upper right), \citet{Schaller1992} (lower left)
and \citet{Meynet2005} (lower right). As our analysis only allows a
limited set of discrete values, many of the points contain multiple
stars.
}
\label{fig:iso}
\end{figure*}

To use this information it is necessary to derive the stellar masses
from the spectral types. We do this separately for the main sequence
stars and for the evolved stars. As in \citet{Voss-orion} we find the 
temperature and luminosity of the stellar spectral types of O stars 
from the line-blanketed models of \citet{Martins2005}, using their 
observational scale. The masses were then found by comparing to
the rotating stellar models of \citet{Meynet2005}.
In \citet{Voss-orion} we compared the masses
found from this method to the spectroscopic masses and found
agreement, and a similar result was found by
\citet{Weidner2010}. Fig. \ref{fig:iso} compares the properties of 
the observed stars to isochrones from 4 different sets of stellar models.
The analysis above was repeated for each of the sets, and the
differences were found to be negligible compared to the uncertainties
of our results.
We used a similar approach for the most massive B-stars, where we
took the extrapolation of
the \citet{Martins2005} luminosities presented in \citet{Smith2006},
together with the effective temperatures of \citet{Zorec2009}. The
luminosities are only available for B-stars of luminosity class
III and V. Luminosities for class IV were found by logarithmic
interpolation between class III and V.
B-stars of luminosity class I and II were assumed to
be near the end of the main sequence, and were not included in
this analysis. These stars were included as evolved stars in the
analysis below.

It is not possible to derive the masses of evolved stars with this
method. Instead we made crude estimates of the initial mass of these
stars based on the ages of the clusters. This was done by assuming that 
the mass of the evolved stars must be above the most massive main 
sequence star in the host cluster and below the mass for which the stellar
lifetime is equal to the age of the cluster. The list of evolved stars
is given in Table \ref{tab:evolved}. Some of the WR stars are not
associated with clusters. We do not attempt to derive their ages, and
the uncertainties resulting from these objects is discussed in Sect. \ref{sect:26al}.

\begin{table*}
\begin{center}
\caption{The evolved stars in the Carina region.}
\label{tab:evolved}
\begin{tabular}{lllrr}
\hline\hline
Name & Type & Cluster & Initial mass& $^{26}$Al\\
&&&$M_{\odot}$ & ($10^{-4}M_{\odot}$)\\
\hline
HD 92809  & WC6 & Bochum 10 & $\sim30$ & 0.3--1.5\\
HD 92964  & B2 Ia & Bochum 10 & $\sim30$ & 0.0--0.3\\
HD 92852  & K1/3 III & Bochum 10 & $\sim30$ & 0.0--0.3\\
Evans 3293-001 & B0 Iab & NGC 3293& $\sim20$ & 0.0--0.1\\
Evans 3293-002 & B0.7 Ib & NGC 3293 &$\sim20$ & 0.0--0.1\\
Feinstein CPD 3502 & M1.5 Iab-Ib & NGC 3293 &$\sim20$ & 0.0--0.1\\
HD 93129Aa  & O2 If\* & Trumpler 14 &80-120 & 0.0--4.0\\
$\eta$ Car & LBV & Trumpler 16 & 80-120 & 0.0--4.0\\
HD 93162 & WN6ha & Trumpler 16 & 80-120 & 2.0--8.0\\
HD 93131 & WN6ha & Trumpler 16 & 80-120 & 2.0--8.0\\
HD 92740A & WN7ha & Trumpler 16 & 80-120 & 2.0--8.0\\
Hucht WR27 & WC6 & Field (2.5)& - & - \\
HD 94546 & WN4 & Field (4.05)& - & -\\
Hucht WR31a & WN11h & Field (8.0)& - & -\\
HD 90657 & WN5 & Field (3.88)& - & -\\
HD 95435 & WC5 & Field (6.11)& - & -\\
Hucht WR21a & WN6 & Field (?)& - & -\\
\hline
\end{tabular}\\
\tablefoot{Stars without HD designations
are labelled by the name of the first author of the publication in which they
appear (see Sect. \ref{sect:stars}, followed by the name given in that 
publication. The initial masses are estimated by the maximum mass of stars
still present in a population with the age of the host cluster. The
ejected amount of $^{26}$Al is found from the stellar tracks at the given
mass. The range is over all WR-phases for WR-stars, and until the onset
of the WR-winds for non-WR stars.}
\end{center}
\end{table*}

\subsection{Initial mass function}
The set of initial masses
can be used to derive an initial mass function (IMF) for the region.
The IMF of Trumpler 16 was previously derived by \citet{Massey1993}
who found it to be consistent with a single power-law with a slope
of $\Gamma$=1.3$\pm0.2$ above 15$M_{\odot}$, where $dN/dM=M^{-\Gamma-1}$
(the Salpeter slope is $\Gamma=1.35$). The stellar content of 
the group has been
revised several times since then, in particular many binary components
have been resolved, and the stellar evolutionary and atmosphere models 
have changed significantly. Furthermore \citet{Massey1993} assumed Gaussian
statistics despite having bins with very few counts. We therefore
provide a new fit to the IMF, fitting a single power-law to
the initial stellar masses using maximum likelihood fitting.

Each of the clusters with a population large enough to achieve
a meaningful fit were fitted individually, assuming that all stars
above 15 $M_{\odot}$ have been identified. The results are given
in Table \ref{tab:IMF}. Clearly they are all consistent with the
\citet{Salpeter1955} mass function, but only Trumpler 16 provides
relatively good statistics. A combined sample consisting of all
the stars in the clusters that samples the complete 15--120 $M_{\odot}$
range yields a very similar result to the fit from Trumpler 16 alone.
We also perform a combined fit to all the clusters, with an upper limit
of 40 $M_{\odot}$ (to avoid incompleteness corrections). Interestingly
this yields a relatively shallow slope. However, the smaller range
means that this is relatively strongly affected by systematic errors
on the stellar masses as well as a possible incompleteness of the
sample at masses $\sim15 M_{\odot}$ and it is not clear that the result
should be trusted. To understand the dependence on these effects we
have performed the fits with a higher completeness mass of $25 M_{\odot}$,
yielding results very similar to the \citet{Salpeter1955} mass function.

\begin{table*}
\begin{center}
\caption{The best-fit IMF of open clusters in our study.}
\label{tab:IMF}
\begin{tabular}{lcccccc}
\hline\hline
Cluster & IMF ($>15M_{\odot}$) & IMF ($>25M_\odot$) & Mass range & No. of stars & Supernovae & Total No. ($>15 M_{\odot}$)\\ 
& $\Gamma^1$ & $\Gamma^1$ &\\
\hline
Bochum 10 & - & - & 15--32 & 5 & 2.34 & 7.34\\
Bochum 11 & $1.8^{+1.1}_{-0.9}$& - & 15--120 & 6 & 0 &6\\
Loden 153& - & - & 15--45 & 2 & 0.43 & 2.43\\
NGC 3293& - & - & 15--22 & 23 & 20.5 & 53.5\\
NGC 3324 & $1.4^{+1.3}_{-1.1}$& - & 15--120 & 3 & 0 & 3\\
Trumpler 14 & $1.3^{+0.6}_{-0.6}$& - & 15--120 & 13 & 0 & 13\\
Trumpler 15 & - & - & 15--40 & 11 & 3.08 & 14.08\\
Trumpler 16 & $1.2^{+0.3}_{-0.3}$& 1.3$^{+0.5}_{-0.5}$ & 15--120 & 54 & 0 & 54\\
All young$^{2}$ & $1.2^{+0.3}_{-0.2}$& $1.2^{+0.5}_{-0.4}$ & 15--120 & 76 & 0 & 76\\
All$^{3}$ & $0.6^{+0.5}_{-0.4}$& $1.3^{+0.4}_{-0.4}$ & - & - & 26.35 & 153.35\\
\hline
\end{tabular}\\
\tablefoot{The upper limits
on the mass ranges were estimated by the age of the host clusters. The number
of stars inside the mass ranges were extrapolated using the Salpeter IMF
to find the number of supernovae and the total number of initial stars.\\
$^{1}$Slope of a single power-law fit.\\
$^{2}$Clusters young enough that no supernova explosions are believed to have occurred.\\
$^{3}$Clusters with maximum stellar masses $\geq 40 M_{\odot}$ fitted in the 15--40 $M_{\odot}$ range.
}
\end{center}
\end{table*}

In a recent survey of Trumpler 15, \citet{Wang2011} found a lack of massive
stars ($>20 M_{\odot}$) by extrapolating observations of stars of lower
masses. Using the \citet{Kroupa2001} mass function, they find that there
should be $\sim11$ stars, whereas they claim that none are observed.
They conclude that either it is an anomaly of the IMF, or alternatively that
all stars above $20M_{\odot}$ have already exploded, which would indicate
an age above 10 Myr, older than what is normally assumed for this cluster.
Contrary to their results, our analysis finds that there are 6 stars in
this cluster with masses above $20M_{\odot}$ and that with an age of $\sim6$
Myr, approximately 3 supernova have exploded, which is in agreement
with the results at lower stellar masses. We believe that the discrepancy
is the result of \citet{Wang2011} underestimating the masses of stars
with early-type spectra. 

\citet{Feinstein1995} summed up the total masses of the observed stars in the
individual clusters. With the updated stellar populations
we improve these estimates. We furthermore take into account the
completeness limits and the contribution from exploded stars to
compare the \textit{initial} stellar content of the individual groups.
Our results show that TR 16 is still the dominant group, but that
the older group ($\sim$10 Myr) NGC 3293, that is often neglected due
to the lack of O stars, has previously hosted a decent population of
these ($\sim10$ stars are expected to have exploded as supernovae in
this association). Assuming the \citet{Salpeter1955} mass function,
the Carina region hosts $\sim3-4$ times as many stars as the more
nearby Orion region \citep{Voss-orion}. This is lower than what is
expected from the observed number of O-stars (a factor of $\sim12$),
due to the somewhat higher age (5--6 Myr) of the bulk of the stars
in Orion.

We furthermore use the \citet{Salpeter1955} IMF to estimate the
relative sizes of the open clusters, by extrapolating the number
of stars in the observed mass ranges of the individual clusters.
Estimates of the initial numbers of stars ($>15 M_{\odot}$) are
given in Table \ref{tab:IMF}, together with the estimated number
of SNe that have already exploded. While the young clusters ($<$3.5 Myr) 
clearly dominate visually (and in terms of how well studied they are),
especially the clusters that are part of the extended Carina Nebula,
they only comprise half of the star formation within the last
$\sim$10 Myr. 

\section{$^{26}$Al}
\label{sect:26al}
The amount of $^{26}$Al found from different analyses of CGRO and
INTEGRAL data is in the range 0.004--0.009$M_{\odot}$.
Even the lower estimates are significantly higher than 
the largest expected mass emitted by a single object. 
The possibility of a much lower mass
ejected by a foreground object was dismissed by \citet{Knoedlseder1996}.
The current understanding that $^{26}$Al is almost exclusively being
ejected by massive stars further limits the possibility of confusion
by foreground objects, due to the completeness of the detection of
nearby massive stars. The background of the Carina complex is less
well understood, but clearly hosts young clusters with massive stars.
For example the massive young cluster Westerlund 2 is within our
field, at a distance of 8 kpc \citep{Rauw2007}. 
However, most of these are at large distances (at least twice the
distance of the Carina Nebula) and within our search radius only
Westerlund 2 has a content of young stars that can compare to Trumpler 16.
The background can therefore only contribute a modest fraction of the
observed signal.

To investigate the origin of the observed $^{26}$Al signal, we divide
the stars into three categories: 
\begin{itemize}
\item main sequence (O) stars
\item exploded stars (SNe)
\item evolved (WR and supergiant) stars
\end{itemize}
and explore their contributions separately. We ignore the population
of stars of spectral type B or later, as their wind contribution is
negligible, and their lifetimes are long enough that they have not
yet gone supernova.

\paragraph{$^{26}$Al from O stars:}
The initial stellar masses and current ages of the 66 O-stars with
spectral types of O3 or later were
estimated above. From  
following the wind ejection and radioactive decay of $^{26}$Al along the  
stellar tracks, we can associate a $^{26}$Al mass to each cluster along  
its evolution.
The sum is found to be 5$\times10^{-5} M_{\odot}$. While
the average ages of the stars are pretty stable using the isochrone
fitting, the individual ages are not very reliable, and the ages of the
few most massive stars are quite important for the result. We therefore
calculate an upper limit to the O-star contribution by assuming that
all the O-stars are just about to evolve off the main sequence (this is
the maximum age of the O-stars as they will change spectral type after
this). This assumption gives an upper limit of 1.5$\times10^{-3} M_{\odot}$.
In this estimate the contribution from the O2 star in Tr 14 as well
as $\eta$ Carinae was ignored. They are included in the discussion of
the evolved stars instead.

\paragraph{$^{26}$Al from supernovae:}
We use our estimate of the number of supernova explosions in each
cluster (see Table \ref{tab:IMF}), together with the population
synthesis tool described in \citet{Voss-popsyn,Voss-orion} to
estimate the contribution from exploded stars to the observed $^{26}$Al.
For each cluster the expected contribution and the error on the estimate
is calculated using monte carlo simulations. In each simulation, the
number of exploded stars is chosen from a Poissonian distribution with the mean
values given in Table \ref{tab:IMF}, and the masses of the stars are
chosen randomly from the \citet{Salpeter1955} mass function. The lower
mass limit of the IMF is given by the stellar track with the lowest
initial mass that has
a lifetime shorter than the current age of the cluster. 
The
ejection of $^{26}$Al from the winds and supernovae is followed
taking into account the radioactive decay. This gives an estimate
of 9$^{+5}_{-4}\times10^{-4} M_{\odot}$ of $^{26}$Al from the supernova yields
and 3$\pm2\times10^{-4} M_{\odot}$ from the preceding winds of the exploded
stars.

\paragraph{$^{26}$Al from evolved stars:}
The final contribution to the $^{26}$Al signal comes from the winds of
the evolved stars. In Table \ref{tab:evolved} we list the evolved stars
that we have identified inside the region. The cluster association of
each evolved stars is listed. A number of WR stars in the catalogue
of \citet{vanderHucht2001} have no cluster association and they are listed
as field sources, with the photometric distances given by the catalogue
(only WR stars with estimated distances below 10 kpc are included in
our list).
The initial masses of the evolved cluster stars were assumed to be
close to the maximum initial stellar mass given the age of the clusters.
For each star we have identified stars in our stellar isochrones with
similar initial masses and evolutionary states (spectral types)
and used these to estimate the possible range of $^{26}$Al masses
present in the ISM.
These are also listed in Table \ref{tab:evolved}.
The field WR stars are more problematic, as their distances are unreliable,
and there is no stellar population through which their approximate age
can be deduced. However, due to the slope of the IMF and the shorter
lifetimes of massive stars, most of them probably correspond to relatively
low (for WR stars) initial masses of (25--40 $M_{\odot}$). As there
is no foreground population of massive stars that they can belong
to, most of them must be located behind the Carina Nebula. This is
consistent with their estimated distances (however uncertain), 
and their contribution is therefore
unlikely to be significant.
From Table \ref{tab:evolved}, it can be seen that the total mass of 
$^{26}$Al from all the evolved stars is unlikely to exceed 
3$\times10^{-3} M_{\odot}$.
The most realistic estimate range is 1.5--2.0$\times10^{-3} M_{\odot}$.

\begin{figure}
\resizebox{\hsize}{!}{\includegraphics[angle=0]{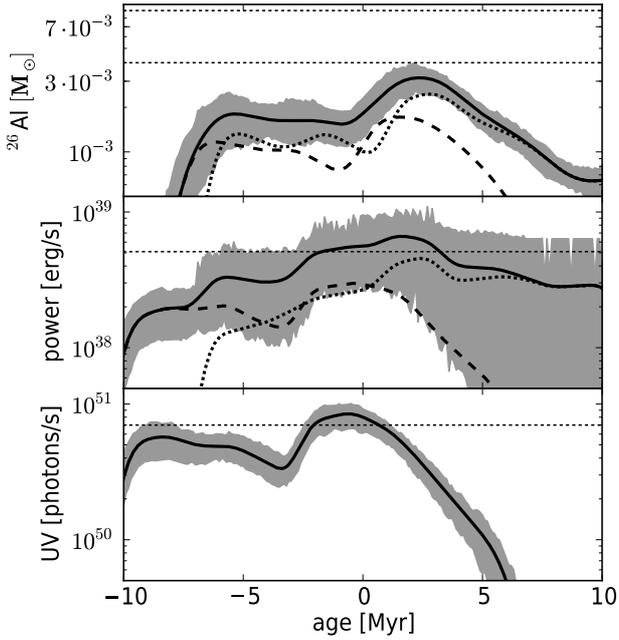}}
\caption{Population synthesis model of the Carina region, for a combined
population of stars from all the stellar clusters with the ages given
in Table \ref{tab:regs} and numbers of stars given in Table 
\ref{tab:IMF}. Time zero corresponds to the current time. The upper
panels shows the mass of $^{26}$Al in the ISM of the Carina region,
the middle panel shows the mechanical power ejected into the
ISM from the massive stars, and the lower panel shows the emitted
flux of hydrogen ionizing photons ($>13.6$ eV). The solid
lines shows the total output from the stellar population, the dashed
line show the output from the winds and the dotted shows the output
from supernova explosions. The grey shaded areas show the 1$\sigma$
deviations of the total output caused by random sampling of the IMF.
The horizontal dashed lines indicate estimates based on
observations, as discussed in the population synthesis part of
Sect. \ref{sect:26al}.
}
\label{fig:26al}
\end{figure}

\paragraph{Population synthesis:}
For comparison we compute the expectation of $^{26}$Al from the
entire region using population synthesis \citep{Voss-popsyn}, with
the cluster ages from Table \ref{tab:regs} and total star-numbers
from Table \ref{tab:IMF}. Each cluster was assumed to have a Gaussian
age spread with $\sigma$=0.5 Myr. We used the model that gives the highest
$^{26}$Al yields, with rotating stellar evolution models from
\citet{Meynet2005} and supernova yields from \citet{Limongi2006}.
The results are shown in the top panel of Fig. \ref{fig:26al}.
Also shown as the range between the two horizontal dashed lines,
is the most likely observed $^{26}$Al signal. As discussed in Sect.
\ref{sect:observations} the range should not be taken as a statistical 
confidence interval, due to the poorly constrained systematic effects of the
different ways to measure the signal. 
The population synthesis model predicts that the signal is divided
roughly evenly between the wind and the supernova contributions.
This is mainly due to the fact that the population synthesis
model predicts a lower wind contribution than what is found from
the observed stars. However, this is consistent within the 90\%
variance caused by the sampling of the population of massive stars
\citep[see discussion in][]{Voss-popsyn}. It is therefore just an
effect of the actual realization of the population of massive stars,
which yields a higher wind contribution to the $^{26}$Al than an
average cluster (with the same properties).
Also the slope of the time-profile is relatively steep, and it is therefore
clear that small errors in the estimates of the cluster ages
(especially of the youngest clusters) can have a relatively large
influence on the conclusions from the population synthesis.

In the middle panel of Fig. \ref{fig:26al} we show the energy injected
into the ISM
from the stars, predicted by the population synthesis. The total output
can be seen to have risen to $\gtrsim2\times10^{38}$ erg s$^{-1}$ about
10 Myr ago, and then slowly increased to the current output of
$\sim5-6\times10^{38}$ erg s$^{-1}$. The energy budget of the associations
in the Carina Complex was studied by \citet{Smith2006,Smith2007}, who
found the stellar wind output to be $\sim2\times10^{38}$ erg s$^{-1}$.
This result did not take supernova contributions into account, and did
not include NGC 3293. Taking these differences into account, their
results are in good agreement with our population synthesis. 
They also estimated the mechanical energy needed to create
the superbubble surrounding the region to be $8\times10^{51}$ erg.
This corresponds to $\sim5\times10^{37}$ erg s$^{-1}$, assuming a
constant power over the last 5 Myr. The horizontal dashed line
in the middle panel of Fig. \ref{fig:26al}
shows this observational estimate. It has been
multiplied by a factor of 10 to account for the observations from
other regions showing that only
a fraction of $\sim10--20$\% of the kinetic energy goes into the
expansion of the bubble \citep[e.g.][]{Brown1995,Cooper2004}.
The height of the line is in good agreement with the expectations,
arguing for a similar energy efficiency of creating the superbubble
in the Carina region.
Our population synthesis predictions for the emission of
hydrogen ionized UV photons are shown in the lower panel of
Fig. \ref{fig:26al}. The UV emission
was also studied
by \citet{Smith2007}. They found a rate
just below $10^{51}$ s$^{-1}$, in good agreement with our population
synthesis model. From observations of the radio continuum, they
deduced that the ISM around the Carina region absorbs 
$\sim7\times10^{50}$ ph s$^{-1}$ of UV radiation, shown as the 
horizontal dashed line in the lower panel of Fig. \ref{fig:26al} 
and the remaining
$\sim25$\% of the flux leak out from the region.

\section{Discussion}
\subsection{Comparison with other young associations}
\label{sect:comparison}
Other regions were investigated recently. Population synthesis models
of the regions Orion \citep{Voss-orion}, Cygnus \citep{Martin2009,Martin2010} 
and Sco-Cen \citep{Diehl2010} were found
to be consistent with observations. Similar to
the results presented above, the analysis of both Orion and Cygnus
supports theoretical models with high $^{26}$Al ejection from the
WR-winds of the massive stars. However, due to the high statistical
fluctuations of the $^{26}$Al output from individual regions, 
caused by the random sampling of the IMF each result is of low 
significance ($<2\sigma$). 

An interesting aspect of comparing different regions is the
differences in their star-formation histories. In Orion and
Sco-Cen the $^{26}$Al output is dominated by populations
of stars with ages $>5$Myr, where the ejection from supernova
explosions is most important. Therefore more than 50\% of the
$^{26}$Al in these regions is expected to be emitted by the
supernovae. Therefore these observations alone could not be used to
distinguish between models with strong wind ejection, as all the
$^{26}$Al could be emitted by supernovae if their yields were
enhanced by a factor of $\lesssim$2.
In the Carina and Cygnus regions the situation is different. In Carina 
we have estimated that only $\sim20$\% of the signal comes from supernovae,
and also in Cygnus the wind contribution dominates 
\citep{Martin2009,Martin2010}. 
An increase of the supernova ejection by a factor of $\sim5$ would be
needed to explain the signal with $^{26}$Al from supernovae alone,
but this is not consistent with the observations of Orion, and Sco-Cen. 
It is therefore clear that the $^{26}$Al signals observed from
Carina and Cygnus
are not due to supernova ejecta alone, and a strong wind ejection
component is necessary. This is therefore the strongest evidence 
of $^{26}$Al ejection in WR-winds to date.

\subsection{Possible explanations for the high $^{26}$Al signal}
\label{sect:tvo}
We have shown above, that the most likely observed $^{26}$Al
signal is higher than the theoretical predictions, even for the
models with the strongest wind ejection.
Here we discuss possible effects that were not taken into account
in our analysis:\\
\paragraph{Hidden stars:}
Recent results have shown that the population of OB stars in the
Carina region has been underestimated by a factor of $\sim50$\%
\citep{Povich2011}. The extra OB stars are found outside the
clusters discussed above with some of them being in recently
identified clusters and the rest being distributed between
them. The reason these stars were not identified before is the
large absorbing column in front of them, caused by molecular
clouds in their local environment. The average visual extinction
of the OB candidates of \citet{Povich2011} is $A_V=5.8$, and the
most absorbed candidates have an extinction $A_V>30$. As $^{26}$Al
is observed with $\gamma$-rays, it is not affected by extinction, and
the hidden population can therefore provide a significant contribution
to the signal. However, we believe that the
hidden population contributes less than expected from the number of OB stars.
The large absorbing columns indicate that many of the stars are still
inside or near their parent molecular clouds. For example the
Treasury Chest cluster is estimated to be younger than 1 Myr
\citep{Smith2005}.
Groups of stars with the age where the $^{26}$Al signal is strongest 
(3--10 Myr) provide enough kinetic energy and UV radiation to ionize/disperse 
nearby clouds, and
are therefore much less likely to be obscured. On the other 
hand some of the OB stars could be located behind other clouds. In the
Cygnus region, this is for example the case of Cygnus OB2, which is
heavily obscured, but contributes significantly to the $^{26}$Al signal
\citep{Martin2009,Martin2010}. In Cygnus the contribution from a
diffuse population of massive stars was estimated to be $\sim33$\%.
From the discussion above we believe that $^{26}$Al from
stars outside clusters and in hidden clusters that were not taken 
into account in our analysis does contribute to the signal, but with 
less than $33$\% (corresponding to the fraction of hidden massive stars) 
of the total signal.\\
\paragraph{Nucleosynthesis uncertainties:}
There are large uncertainties in the modelling of supernova 
explosions \citep{Woosley1995,Limongi2006}. Furthermore the
uncertainties in the
nuclear reaction rates responsible for the formation of $^{26}$Al ejected
in the supernova explosions lead to uncertainties of a factor $\sim3$
\citep{Iliadis2011}. Both could explain the
results, if only the Carina region was observed. However, as discussed
above, the observations of the Sco-Cen and Orion regions
contradict strongly enhanced supernova yields. The nuclear
reaction rates responsible for the production of $^{26}$Al ejected
in stellar winds are much better constrained \citep{Iliadis2011},
and the strong signal can therefore not be explained by the
uncertainties of these.\\
\paragraph{Binaries:}
The yields from close binaries are challenging to quantify
\citep{Langer1998}. Such systems experience mass transfer episodes, 
which can lead to
enhanced mass loss and mixing and early exposure and ejection of 
chemically enriched layers of the stars. Furthermore, tidal forces 
affect their rotation and mixing. This can in principle lead
to very high enhancement factors. However, the enhancement in binaries is
concentrated in very few particular systems. Large enhancements in
a few systems would lead to high fluctuations in the $^{26}$Al signal
over the sky, which are not seen. We therefore conclude that while
enhanced systems might explain a modest difference between the
theoretical models and the observations, most of the $^{26}$Al is
being ejected by ``normal'' stars.\\
\paragraph{Very massive stars:}
There is growing evidence that
stars can initially be more massive than the limit of 120 $M_{\odot}$
assumed in our study \citep{Figer2005,Koen2006,Crowther2010,Bestenlehner2011}, 
and the fact that many stars are born in multiple
systems can both increase the yields significantly. Indeed, both
have been invoked to explain properties of stars in the Carina Nebula,
in particular $\eta$ Carinae.
The $^{26}$Al yields of stars with masses above 120 $M_{\odot}$ have
not been studied. However, the yields of stellar models increase
relatively strongly with initial mass, and it is therefore likely
that initially very massive stars can yield several times the amount
of $^{26}$Al given by our most massive stellar model.
Similar to the binaries, Large enhancements in
a few stars would lead to high fluctuations in the $^{26}$Al signal
over the sky, which are not seen, and we therefore believe that the
possible contribution from such very massive stars must be modest.

\subsection{Implications for stellar evolution}
The effects discussed above are likely to be minor. It
is possible that a combination of these effects are enough to
account for the observed $^{26}$Al signal being higher than the
highest theoretical model, including both a large wind and a
large supernova contribution. We do therefore not find that
the apparent discrepancy calls for significant changes in the
modelling of massive stars. However, only $\sim20$\% of the
signal can be explained by $^{26}$Al ejected by supernova explosions.
As discussed above in Sect. \ref{sect:comparison}, the observations
of other regions do not allow the supernova yields to be raised by
a factor of few, and our results are therefore in support of a
strong wind ejection.

Despite the downwards revision of the mass loss rates in the latest
stellar evolutionary models \citep{Vink2000,Meynet2005}, the integrated mass
loss has not decreased significantly, and the $^{26}$Al ejection has
actually increased \citep{Palacios2005}. This is due to the effects of
rotation that makes the stars spend longer time
in states with high mass loss rates, and the meridional circulation is
increased, lifting $^{26}$Al to the surface. A further downwards revision of
the wind mass loss rates \citep[as suggested by e.g.][]{Fullerton2006},
would reduce the wind yields significantly, and our analysis might
therefore be evidence against such a downwards revision \citep[see also
the discussion in][]{Voss-orion}. Similarly, non-rotating models
with current wind mass-loss prescriptions have low wind ejection of
$^{26}$Al, and our results are therefore in support of rotational
effects being significant.

\subsection{Conclusions}
We have studied the population of massive stars in the Carina region.
Our analysis of the ejection of kinetic energy into the ISM and
the emission of ionizing UV radiation, and our results are in
agreement with previous theoretical estimates and observational results.
$^{26}$Al is an important tracer of massive star evolution and 
the interaction between massive stars and their surroundings. 
We have constrained the $^{26}$Al signal 
from the region around the Carina Nebula, and shown that it is 
consistent with coming from the populations of massive stars in this region.
Our results show that most $\sim80$\% of the $^{26}$Al was ejected
by the winds of massive stars. This result strongly favours rotating
stellar evolutionary models, and is in disagreement with the suggested
further reductions of the mass-loss rates due to clumping beyond what
is included in the latest generation of models.

\begin{acknowledgements}
This research is supported by NWO Vidi grant 016.093.305.
PM acknowledges support from the European Community via
contract ERC-StG-200911. Based on observations with INTEGRAL, 
an ESA project with instruments 
and science data centre funded by ESA member states. 
\end{acknowledgements}

\end{document}